\documentclass[journal=aamick,manuscript=article,layout=traditional]{achemso}
\usepackage{achemso} \usepackage{float}
\usepackage{gensymb}
\usepackage{amsmath}
\usepackage{color}
\setkeys{acs}{articletitle=true}
\setkeys{acs}{maxauthors = 0}

\author{Boxiao Cao} \affiliation{Department of Civil and Environmental
  Engineering, George Washington University, Washington, DC 20052}
\author{Shunda Chen} \affiliation{Department of Civil and
  Environmental Engineering, George Washington University, Washington,
  DC 20052}
\author{Xiaochen Jin} \affiliation{Department of Civil and
  Environmental Engineering, George Washington University, Washington,
  DC 20052}
\author{Jifeng Liu} \affiliation{Thayer School of Engineering,
  Dartmouth College, Hanover, NH 03755}
\author{Tianshu Li} \affiliation{Department of Civil and
  Environmental Engineering, George Washington University, Washington,
  DC 20052} \email{tsli@gwu.edu}

\title{Short-range order in GeSn alloy}


\begin{document}

\begin{abstract}

Group IV alloys have been long viewed as homogeneous
random solid solutions since they were first perceived as Si-compatible,
direct-band-gap semiconductors 30 years ago. Such a perception underlies the understanding, interpretation
and prediction of alloys' properties. However, as the race to create
scalable and tunable device materials enters a composition domain far beyond alloys' equilibrium solubility,
a fundamental question emerges as to how random these alloys truly are.      
Here we show, by combining statistical sampling and large-scale
\textit{ab initio} calculations, that GeSn alloy, a promising group IV
alloy for mid-infrared technology, exhibits a clear, short-range order
for solute atoms within its entire composition range. Such short-range
order is further found to substantially affect the electronic
properties of GeSn. We demonstrate the proper inclusion of this short-range
order through canonical sampling can lead to a significant
improvement over previous predictions on alloy's band gaps, by showing
an excellent agreement with experiments within the entire studied composition range.  
Our finding thus not only calls for an important revision of
current structural model for group IV alloy, but also suggests
short-range order may generically exist in different types of alloys. 

\end{abstract}

\section{Introduction}

Since the first prediction in SiGeSn alloy on its direct band gap
\cite{Soref:1991ce}, group IV alloys have attracted 
attentions for silicon photonics. This interest has been significantly
boosted recently as group IV alloys have been indeed
demonstrated viable for optoelectronic applications when a substantial
amount of Sn is incorporated into Si and Ge by physical or chemical
vapor deposition techniques \cite{Gencarelli:2013go,Wirths:2016ku}. In
particular, with a Sn composition beyond 8 at. \%, GeSn alloy was
characterized as a direct band gap semiconductor by photoluminescence
and lasing \cite{Ghetmiri:2014hh,
  Wirths:2015dh,Stange:2016ip,Reboud:2017cr,AlKabi:2016de,
  Margetis:2017fs,Li:2018ka}. As the band gap of GeSn alloy
continuously decreases with Sn composition, a high-Sn content GeSn
alloy ({\em e.g.}, $>$20 at.\% Sn) is of particular interest as it enables a
wavelength coverage well within the mid-infrared range, thus promising
numerous exciting applications including light-emission, chemical and
biological sensing \cite{Soref:2010iw}.

One of the main challenges in GeSn alloy is Sn segregation, where Sn
atoms aggregate in lattice, adversely
affecting the properties and integrity of GeSn alloy. To address this issue,
experimental characterization has been focused on differentiating Sn
segregation from Sn dispersion \cite{Kumar:2013ih,Kumar:2015ha,Assali:2017et,Assali:2018db}. As a commonly adopted notion, GeSn alloy is viewed as a homogeneous random solid
solution when no Sn segregation is present. This assumption has been employed to
interpret experiments \cite{Kumar:2013ih,Kumar:2015ha,Assali:2017et,Assali:2018db,Xu:2019bo,Doherty:2020id}, but more crucially,
constitutes the foundation for nearly all theoretical predictions of
GeSn alloys \cite{Shen:1997fc,Pandey:1999ii,Moontragoon:2007ho,Yin:2008cf,Beeler:2011be,Low:2012bh,Lee:2013ib,Gupta:2013kp,Eckhardt:2014gz,Freitas:2016kc,Polak:2017fh,OHalloran:2019ik}.
Indeed, commonly employed modeling methods, including virtual crystal
approximation (VCA), coherent-potential approximation (CPA), and
special quasi-random structure (SQS) \cite{Zunger:1990tw}, are all
based on this assumption.

Although the absence of Sn
cluster certainly indicates atoms are well dispersed within the
lattice, it remains unclear whether a well-dispersed Sn
distribution necessarily implies a truly random distribution. In a truly
random alloy, a lattice site is occupied by constituent elements with
a probability solely depending on alloy's composition,
irrespective of what is present in its neighbors. In contrast, in a
well-dispersed, but not necessarily truly random alloy, the probability of
site occupancy also depends on neighboring atoms, due to a correlation
between constituent elements. Both types of distributions lead to a
microscopically homogeneous and well-dispersed alloy, {\em i.e.,} no
segregation, but are distinguished by whether a short-range ordering
is present.    

Indeed, recent experimental studies
\cite{Gencarelli:2015fl,Robouch:2020ib} based on extended X-ray
absorption fine structure (EXAFS) already showed a lack of Sn-Sn
nearest neighbors in GeSn alloy with a Sn content up to 12.4
at. \%. Characterization of non-equilibrium SiGeSn ternary alloy by
atom probe tomography also showed an unexpected repulsive interaction
between Si and Sn \cite{Mukherjee:2017is}. On the other hand, although
theory and experiment are found to agree well for low-Sn content GeSn
alloy \cite{Doherty:2020id}, a significant discrepancy emerges very
recently for high-Sn alloys on their fundamental band gaps. In
particular, first principle calculations based on a random solution
model all suggest GeSn alloys turn into a metal at $25\sim 28$ at. \%
Sn content \cite{Eckhardt:2014gz,Freitas:2016kc,Polak:2017fh}, whereas
very recent optical studies confirm ultra-high-Sn content GeSn alloy
still remains as a direct band gap semiconductor
\cite{Xu:2019ds,Xu:2019bo}, for example, a $E_g^{\Gamma}$=0.15 eV for
a Sn concentration as high as 32 at. \% \cite{Xu:2019bo}. As recent
advance in synthesis enables the growth of high-Sn content alloys far
beyond Sn's solubility in Ge and Si
\cite{Reboud:2017cr,Margetis:2017fs,Li:2018ka,Dou:2018ik,Imbrenda:2018fw,Xu:2019ds,Xu:2019bo},
a fundamental question emerges as to how random these group IV alloys
truly are.

Here we examine the fundamental assumption of random solid solution in GeSn
alloys by combining statistical
sampling based on Monte Carlo method and density functional theory
(DFT) calculation. Our study indeed shows that GeSn alloy, in contrast
to SiGe alloy and the conventional notion, is clearly not a truly random
solid solution, by exhibiting a partial, short-range order (SRO) that becomes
particularly prominent in the correlation function involving solute
atoms. This SRO is reflected by a lower-than-random solute-solute
coordination number in their first coordination shell, and resembles what has been recently observed in metallic alloys
\cite{Zhang:2017bw,Ding:2018im,Zhang:2020kr}. Importantly, the
identified SRO is further found to account for the
discrepancy between theory and experiment, through our
ensemble-averaged prediction that explicitly incorporates the
SRO, and shows an excellent
agreement with the experiments over the entire composition range studied so
far. 


\section{Methods}

\subsection{Metropolis Monte Carlo Sampling}

The notion of simple arithmetic average is commonly employed to obtain
the average property of interest $\bar X$ in alloy, namely, $\bar
X=1/N\sum_i^N X_i$, where $N$ is the number of configurations and
$X_i$ is the property of configuration $i$. An underlying assumption
for simple average is that all configurations within the ensemble
carry the same statistical weight, which is a condition satisfied in
truly random solution. In real alloy, each configuration may carry a unique
statistical weight $w_i$, thus contributing differently to the
ensemble average $\langle X \rangle=\sum_i w_i X_i$. Under a given
temperature $T$, such statistical weight is the Boltzmann factor
$w_i=\exp(-E_i/k_BT)/Z$, where $E_i$ is the total energy for the configuration $i$, $k_B$ is
the Boltzmann constant, and $Z$ is the partition function for
canonical ensemble. An efficient way of computing the ensemble average
$\langle X \rangle$ is Metropolis Monte Carlo (MC) method
\cite{Metropolis:1953in}, where a new configuration $j$, created by a
trial move from the current configuration $i$, is accepted based on
$w_j/w_i=\exp(-(E_j-E_i)/k_BT)$. For binary group IV alloy, the trial move
involves swapping a randomly selected solute atom with a randomly selected
solvent atom within Diamond Cubic (DC) lattice. The new configuration is
then fully relaxed by DFT calculation to obtain the total energy $E$.

\subsection{DFT calculations}

Our DFT calculation is carried out using the Vienna Ab initio
Simulation Package (VASP) \cite{Kresse:1993ty} based on the projector
augmented wave method
\cite{Kresse:1999tq,Kresse:1996kg,Kresse:1996vf}. Local density
approximation (LDA) \cite{Ceperley:1980gc} is employed, as previous
studies
\cite{Haas:2009be,Eckhardt:2014gz,Polak:2017fh,OHalloran:2019ik}
showed LDA yields the best agreement with experiment on pure Ge
and Sn for geometry optimization. We also test
Perdew-Burke-Ernzerhof (PBE) functional \cite{PERDEW:1996ug}, to
ensure the results are robust against the choice of
exchange-correlation functionals. Supercells containing 32, 64, and
128 atoms are used to ensure the results are size- and shape
independent. A 64-atom supercell is obtained by replicating a
conventional DC cell containing 8 atoms twice along each dimension,
{\em i.e.}, $2\times2\times2$, whereas a 32-atom supercell corresponds
to a $2\times2\times1$ cell. A 128-atom cell is constructed based on
the primitive cell of DC structure (containing 2 atoms) by repeating
it 4 times along each dimension, {\em i.e.},
$4\times4\times4$. Because the difference of total energy between two
configurations $\Delta E=E_j-E_i$ is of the central importance for
sampling, and $\Delta E$ converges much faster than $E$ itself, only
Gamma point is used to sample the first Brillouin Zone to enhance
computational efficiency. A plane-wave basis with a cutoff energy of
400 eV is used throughout the calculation. For each new configuration,
the supercell undergoes full relaxation where both cell geometry and
atomic positions are relaxed by conjugate gradient algorithm, with the
convergence criteria being $10^{-4}$ eV and $10^{-3}$ eV for
electronic and ionic relaxations, respectively.

\subsection{Calculation of radial distribution function}

Radial distribution function (RDF) is calculated by randomly choosing
$300\sim 600$ snapshots from the obtained MC trajectory based on
64-atom cell,
excluding the first 500 configurations due to equilibration. To take
thermal motion of atoms into account, {\em ab initio} molecular
dynamics (AIMD) at
300~K is carried out on each configuration for 1 ps, yielding a
trajectory of $300\sim 600$ ps from which RDF is calculated. For
obtaining RDF of a truly
random alloy, a comparable number of configurations are randomly
generated, and each undergoes full ionic relaxation, followed by the same AIMD
simulation.

\subsection{Ensemble-averaged band structure calculation}

For band structure
calculations, we employ the modified Becke-Johnson (mBJ) exchange
potential \cite{Tran:2009kk} that has been demonstrated to predict the
correct band gaps of both Ge and $\alpha-$Sn
\cite{Eckhardt:2014gz,Polak:2017fh} with the $c$-mBJ parameter
set to be 1.2. To recover the Bloch character of electronic
eigenstates perturbed by disorder, we apply the spectral weight
approach \cite{Popescu:2010jd,Rubel:2014fv} through the code {\em
  fold2bloch} \cite{Rubel:2014fv} to unfold the band structures back into the first
primitive Brillouin Zone of DC structure. Since the spectral weight
approach requires a supercell generated by a translation of primitive
cell in real space, band structure calculations are carried out
on a 128-atom supercell. To fulfill this requirement, we re-generate
the MC trajectory using a 128-atom cell, which, while expensive,
also serves as a cross validation of our results against size
effect. Spin-orbit coupling (SOC) is also considered in the
band structure calculation. Although SOC was demonstrated crucial for
reproducing the band structures of Ge and $\alpha-$Sn \cite{Eckhardt:2014gz,Polak:2017fh}, it significantly
increases the computational cost, and more importantly, including SOC
is found to lead to a rigid shift in direct band gap, virtually independent of
configurations (see Supporting Information S2). Therefore, for the
part of our study for understanding the role of canonical sampling, we
neglect the SOC in the band gap predictions, whereas SOC is included
when predicting the concentration dependence of direct band gap.       

\section{Results}

\subsection{Short-range order in solute atoms}

Since each atom within a DC lattice is surrounded by four nearest
neighbors, the coordination number (CN), which is defined as the
number of the nearest neighbors of an atom, is four in elemental
semiconductors. If a DC lattice is randomly occupied by element $A$
and $B$, {\em i.e.}, forming a random solution, then the $A-A$ or
$B-B$ coordination number is solely determined by the composition of
the alloy. For example, a Ge$_{0.75}$Sn$_{0.25}$ random solution should
yield a Sn-Sn CN of one, since one of the four nearest neighbors of
each Sn atom, on average, is occupied by a Sn atom, which is set by
the overall composition of Sn ($=0.25$). To examine whether this is
true in group IV alloy, we compute the solute-solute coordination
number in both GeSn (Sn-Sn) and SiGe (Ge-Ge) by integrating the
calculated radial distribution function $g(r)$ based on the obtained
MC trajectory.

\begin{figure}[t]
  \includegraphics[width=0.75\linewidth]{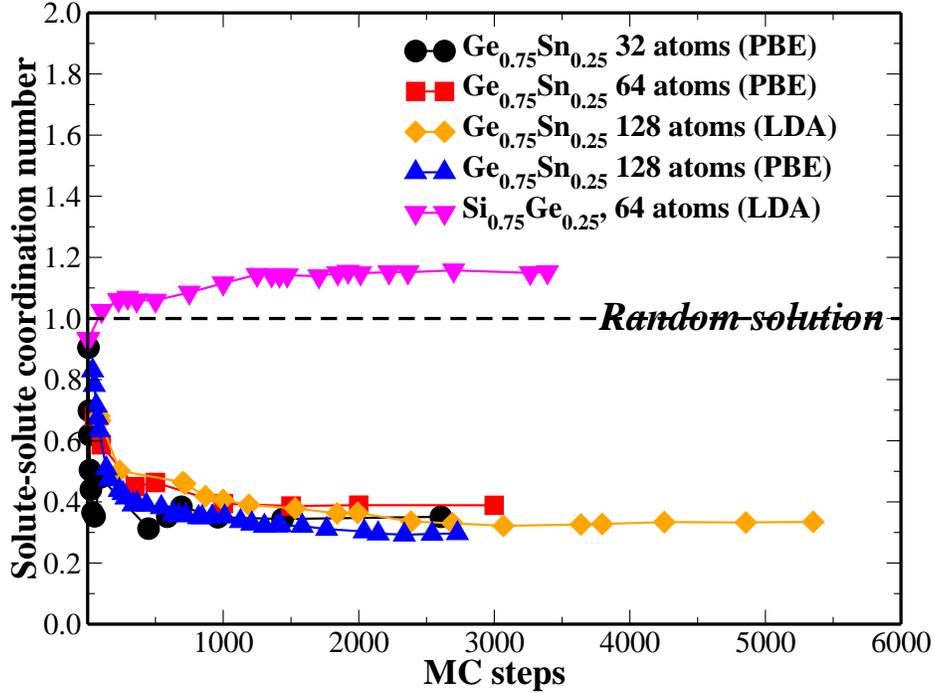}
\caption{\label{CN} Convergence of the calculated ensemble-averaged
  solute-solute coordination number in Si$_{0.75}$Ge$_{0.25}$ and
  Ge$_{0.75}$Sn$_{0.25}$ alloys at 300~K, based on the combined MC/DFT
  method. The dash line indicates the value of solute-solute
  coordination number in a random solution of the same composition. }
\end{figure}

Fig. \ref{CN} shows the variation of the ensemble-averaged
solute-solute coordination number with the number of Monte Carlo steps
in both Ge$_{0.75}$Sn$_{0.25}$ and Si$_{0.75}$Ge$_{0.25}$ alloys at
300~K. It can be seen that beyond $\sim 1,500$ MC steps, the
ensemble-averaged CN has reached a plateau, indicating a numerical
convergence. The converged Sn-Sn CNs are further found to be
independent of the choice of exchange-correlation functional, or
size/shape of the simulation cells, thus confirming the robustness of the
results. Interestingly, the ensemble-averaged Ge-Ge coordination
number in Si$_{0.75}$Ge$_{0.25}$ alloy is found to be around $1.1$,
which is just slightly higher than the ideal value (one) in a truly
random Si$_{0.75}$Ge$_{0.25}$ alloy. This close agreement thus
supports the applicability of the random solution model in SiGe alloy,
and also confirms the effectiveness of the combined MC/DFT method.

In sharp contrast to SiGe alloy, our calculation shows the
ensemble-averaged Sn-Sn coordination number in Ge$_{0.75}$Sn$_{0.25}$
is only $0.33\pm0.02$ at 300~K, namely, one third of that for a random
solution. The significantly lower Sn-Sn coordination number means that
there is a depletion of Sn atoms within the first coordination shell
of a Sn atom in GeSn alloy, or in other words, there is a strong
tendency for a Sn atom to avoid another Sn atom in its first nearest
neighbor. Because each Sn atom is four-fold coordinated, the depleted
Sn-Sn bonds ($\sim 2/3$ for each Sn atom) must be compensated by the
same number of Sn-Ge bonds. This behavior then clearly indicates that
Ge$_{0.75}$Sn$_{0.25}$ at 300~K deviates from a truly random alloy,
{\em i.e.}, Sn atoms exhibiting some degree of SRO
within the alloy. Since the calculated Sn-Sn coordination number is
non zero, the identified SRO should be characterized as
partial.

\begin{figure*}
\includegraphics[width=\linewidth]{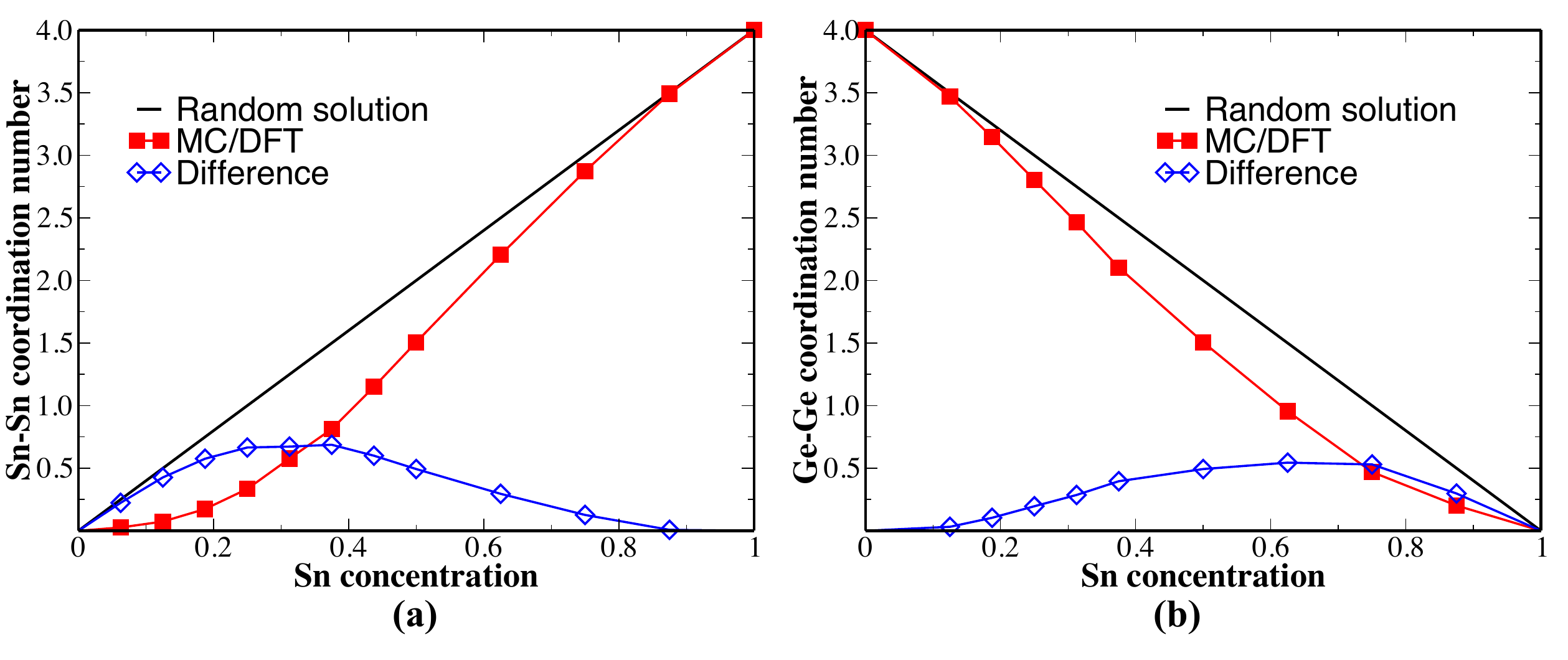}
\caption{\label{CNcomp} Variation of the ensemble-averaged (a) Sn-Sn
  coordination number and, (b) Ge-Ge coordination number with Sn
  concentration in GeSn alloy at 300~K. The error bar of the
  calculated solute-solute coordination number is about $5\sim 10$\%
  of the mean (see Supporting Information S1 for details), which is
  about the size of the symbols.}
\end{figure*}

To understand whether such SRO is unique in
Ge$_{0.75}$Sn$_{0.25}$, or rather generic in all GeSn alloys
regardless of its composition, we carry out extensive calculations of
Sn-Sn CN within the entire compositional range. As shown in
Fig. \ref{CNcomp}, GeSn alloys of different compositions are all found
to exhibit similar behaviors, albeit that the degree of partial
ordering varies with composition. In particular, the deviation from
the ideal Sn-Sn coordination number is found to increase quickly with
Sn concentration for low-Sn alloy, reach the maximum at a Sn content
of around 30 at. \%, and then slowly decrease to zero for pure $\alpha-$Sn.

Remarkably, we find that the SRO is not just limited to
Sn-Sn nearest neighbors, but also exists in Ge-Ge configurations in
GeSn alloy. As shown in Fig. \ref{CNcomp}(b), the calculated Ge-Ge
coordination number also displays a deviation from its ideal
value, and such deviation is also found to be compositionally
dependent. Intriguingly, the composition-dependence in Ge-Ge CN is
nearly symmetric to that in Sn-Sn CN: In Ge-rich alloy, the Sn-Sn
coordination exhibits a fairly strong SRO while the Ge-Ge
distribution stays virtually random; In Sn-rich alloy, the Ge-Ge
coordination becomes less random while the Sn-Sn coordination remains
nearly random. This contrasting behavior thus indicates that the
SRO becomes particularly prominent in the correlation
function involving solute or minority atoms in GeSn alloy.

\begin{figure*}

  \includegraphics[width=\linewidth]{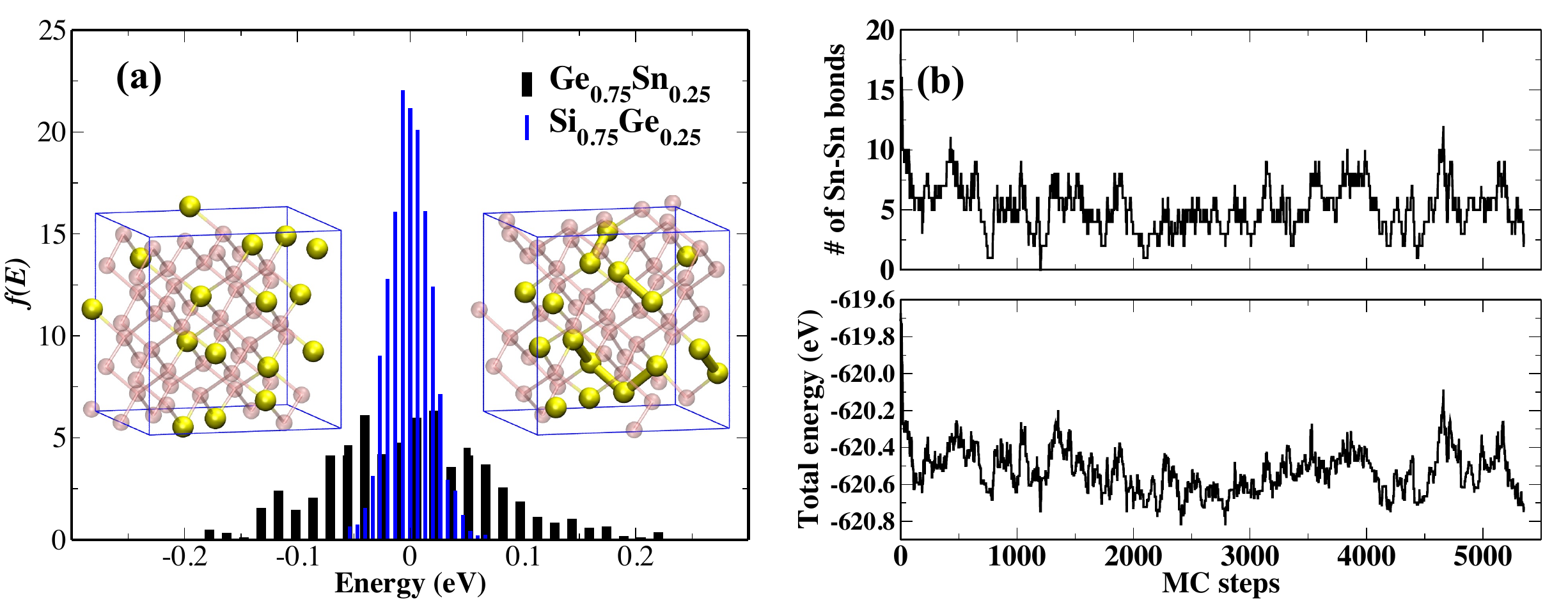}
\caption{\label{dist} (a) Distribution of total energy $E$ within the
  canonical ensembles obtained by MC/DFT calculations for
  Ge$_{0.75}$Sn$_{0.25}$ alloy (black) and Si$_{0.75}$Ge$_{0.25}$
  alloy (blue) at 300~K based on a 64-atoms cell. The mean energies
  $\langle E \rangle$ for both distributions are set to be zero for
  the purpose of comparison. Insets show a low-energy configuration on
  the left and a high-energy configuration on the right. Sn and Ge
  atoms are represented by yellow and pink spheres, respectively. (b)
  Variation of the number of Sn-Sn bonds (upper panel) and the total
  energy (lower panel) in MC trajectory, obtained for
  Ge$_{0.75}$Sn$_{0.25}$ at 300~K based on a 128-atom cell.}
\end{figure*}

To understand the origin of this non-random solution behavior in GeSn alloy, we
examine the distribution of the total energy $E$ within the ensemble
of configurations, as shown in Fig. \ref{dist}(a). In
Ge$_{0.75}$Sn$_{0.25}$ alloy, the analysis finds a relatively wide
range of distribution of $E$ for different configurations: from 0.2 eV
below to 0.2 eV above the mean energy of the ensemble $\langle
E\rangle$. As a comparison, the distribution of $E$ in SiGe alloy is
found to be significantly narrower. A wide distribution of $E$
indicates that a change of atomic configuration in alloy will likely
lead to a structure that is energetically very distinct from the
original configuration. If the new configuration is strongly
unfavorable energetically, it is then less likely to occur within the
ensemble, thus having a negligible statistical weight in ensemble
average. Toward this end, we examine the structures of
Ge$_{0.75}$Sn$_{0.25}$ alloys located in the different regions of
energy distribution, and find that high-energy structures are
generally accompanied by a larger number of Sn-Sn pairs, and sometimes
even small Sn clusters composed of more than two Sn atoms (see the
inset of Fig. \ref{dist}(a)), whereas in low-energy structures, Sn
atoms are largely well dispersed within the Ge matrix, thus avoiding
having other Sn atoms as their immediate neighbors. To confirm this
observation is of general relevance, we compute the number of Sn-Sn
bonds (within a cutoff distance of 3.0~\AA) for each configuration
within the MC ensemble obtained using 128 atoms for
Ge$_{0.75}$Sn$_{0.25}$ alloy. Fig. \ref{dist}(b) shows the variation
of Sn-Sn bond numbers indeed follows closely the variation of the
total energy, suggesting that the nearest neighbor Sn-Sn
configurations play a key role in the non-random solution behavior of
Ge$_{0.75}$Sn$_{0.25}$ alloy.

A natural question is then why SRO is prominent in
GeSn but not in SiGe alloys, given that both are group IV alloys. A
possible explanation can be related to the size difference in group IV
elements. As the lattice constants increase slowly from Si (5.43~\AA)
to Ge (5.66~\AA), but rapidly from Ge to $\alpha-$Sn (6.46~\AA), the
lattice mismatch between Ge and $\alpha-$Sn (14.1\%) is significantly
higher than that between Si and Ge (4.2\%).  Therefore substituting Ge
atoms by larger Sn atoms in the DC lattice of Ge will lead to
substantial local distortions, which are expected to be higher than
those in composition-equivalent SiGe alloys. In particular, such local
distortions will become more significant when Sn atoms are located
adjacent to each other, which then leads to configurations that are
energetically less favorable than those where Sn atoms are well
dispersed within the Ge matrix. Indeed, previous theoretical study
showed Sn-Sn interactions are repulsive when Sn atoms serve as
substitutional defects in Ge lattice \cite{Fuhr:2013fo}. Similarly, in Sn-rich alloys,
substituting Sn atoms in the $\alpha-$Sn lattice by smaller Ge atoms
will also lead to distortions which become significant if Ge atoms are
clustered. This explains why SRO is particularly notable
in the solute-solute, rather than solvent-solvent, coordination number in
GeSn alloys.

\subsubsection{Comparison with experiments}
Experimental verification of SRO in alloy has been proven challenging,
particularly for diffraction-based method, because diffraction
contrast arising from local distortion induced by SRO is inherently
weak as compared to matrix lattice diffraction
\cite{Zhang:2020kr}. Nevertheless, there exist
  experimental characterizations on Sn distribution in GeSn
  alloy, albeit that these characterization studies focused on the
  subject of Sn segregation or dispersion. Specifically, two types of characterization
  techniques have been employed: extended X-ray absorption fine
  structure (EXAFS) which allows probing local environment of atoms,
  and atom probe tomography (APT) that offers information of atomic
  positions in 3D. Here we briefly compare our model with experimental
  results.

\begin{figure*}
  \includegraphics[width=0.75\linewidth]{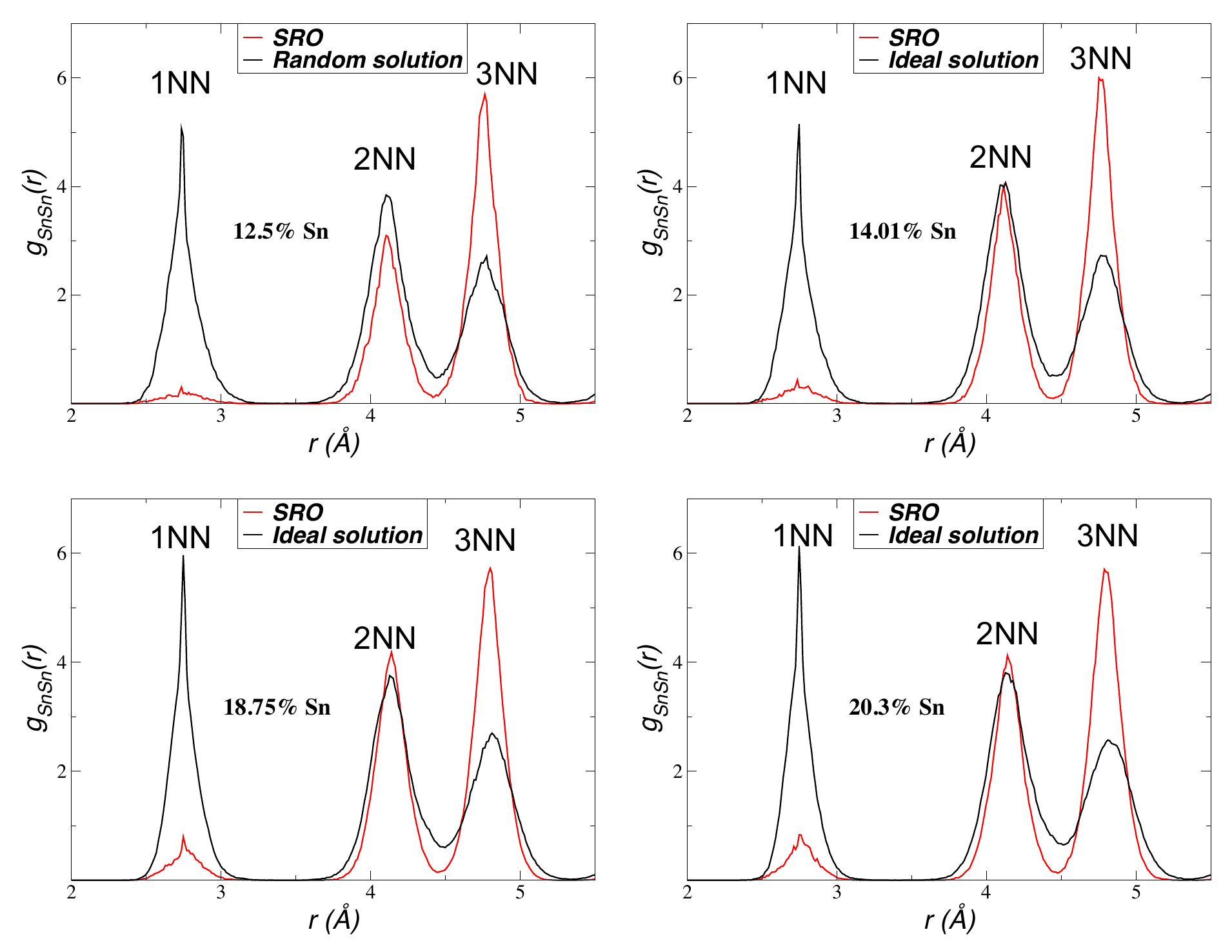}

  \caption{\label{RDF}
    Calculated Sn-Sn radial distribution function of
  GeSn alloys with SRO (red) and without SRO (black) for different Sn
  compositions.}
\end{figure*}

EXAFS studies of Sn local environment in GeSn films \cite{Gencarelli:2015fl,Robouch:2020ib}
provided the experimental evidence in consistent with the type of SRO
in our model.  In particular, the fitted Sn-Sn distance based on EXAFS
data \cite{Gencarelli:2015fl} in GeSn alloy was found to be close to 4~\AA, {\em
  i.e.}, corresponding to the $2^{\text{nd}}$ nearest neighbor (2NN),
rather than the $1^{\text{st}}$ nearest neighbor (1NN) ($\sim 2.8$~\AA),
of Sn in DC lattice. This analysis thus concluded the absence
of Sn-Sn dimers or Sn clusters and also indicated the preference of 2NN Sn
distribution, which is consistent with the identified SRO in our study
where Sn atoms
tend to avoid each other in their 1NN. Indeed, as shown in the calculated Sn-Sn radial
distribution function (Fig. \ref{RDF}), the first Sn-Sn peak in SRO
GeSn is significantly
reduced due to a Sn-Sn repulsion, yielding a peak height much lower
than that of the second Sn-Sn peak. In contrast, in a truly random GeSn alloy, Sn-Sn peak
height decreases with Sn-Sn distance, leading to a peak order opposite to
that of an SRO alloy. We note that this drastic difference
in relative peak height can be an important clue for further
experimental validation. Interestingly, Ref. \cite{Gencarelli:2015fl} concluded GeSn
alloys are
homogeneous random substitutional alloy, on the basis of a relative increase
of Sn atoms in the 2NN of Sn atoms and the absence of Sn cluster. In light of the calculated Sn-Sn
RDF, we note that SRO GeSn alloy can be actually indistinguishable from a
truly random alloy, if an analysis is carried out only based on 2NN Sn,
because as shown in Fig. \ref{RDF}, the second Sn-Sn
peak remains virtually unaffected by SRO. In fact, Fig. \ref{RDF} shows
the depletion of Sn atoms
within the 1NN is largely compensated by the 3NN, thus leaving 2NN
largely intact.

On the other hand, a few APT characterization studies \cite{Kumar:2013ih,Kumar:2015ha,Assali:2017et,Assali:2017et,Assali:2018db} were carried out on chemical vapor
deposition (CVD) grown GeSn films and nanowires, and showed no Sn
segregation or Sn clusters in GeSn alloys. We note that although these studies
indicate a well dispersed Sn distribution, the data need to be refined
to enable further inferring the occurrence of Sn-Sn repulsion in their nearest
neighbors. This is because both a SRO-alloy with a low Sn-Sn
coordination number and a truly random alloy lead to a well dispersion
of Sn atoms, {\em i.e.}, no Sn segregation, within
Ge lattice. In fact, a repulsive Sn-Sn interaction can even facilitate a
better Sn dispersion, as there is a lower chance to have a Sn-Sn
nearest neighbor in SRO-alloy than in truly random alloy. Therefore a
validation of SRO of Sn-Sn repulsion should be obtained by
an explicit comparison of Sn-Sn RDF between a truly random model and a SRO model,
as suggested in Fig. \ref{RDF}. We also note that a validation of SRO
through APT may need a further data processing, analysis, and refinement. For
example, the combination of limited detector
efficiency and imperfect spatial resolution in APT tends to make data
more randomized, as demonstrated in the test of identifying SRO in perfectly
ordered compound \cite{Marceau:2015go}. In addition, the ablation process of APT in peeling off atoms one-by-one may
induce perturbation of atomic positions, which can lead to an
artificial atomic distance. Furthermore, as shown in our subsequent
study \cite{Liu:2020}, SRO itself may exhibit a spatial
heterogeneity due to strain and composition gradient and a temperature
dependence, and depending on the region of samples where APT is
performed, SRO may be difficult to probe or even absent. The detailed discussion on
validating SRO based on APT will be reported in our subsequent
publication \cite{Liu:2020}.

\subsection{Effect of SRO on electronic band gaps}

An important question is how the identified SRO
affects the properties of GeSn alloy, considering nearly all existing theoretical studies
\cite{Shen:1997fc,Pandey:1999ii,Moontragoon:2007ho,Yin:2008cf,Beeler:2011be,Low:2012bh,Lee:2013ib,Gupta:2013kp,Eckhardt:2014gz,Polak:2017fh,OHalloran:2019ik}
were carried out either by simple arithmetic average of randomly
generated simulation cells, or by SQS that matches the correlation
function of a truly random alloy. To answer this question, we calculate the direct band
gap of GeSn alloy, which is a key property for mid-infrared
applications. In order to understand the role of SRO, we explicitly compare the ensemble-averaged band
gaps $\langle E_g^{\Gamma}\rangle$ (canonically sampled) and the
simply-averaged band gaps $\overline{E_g^{\Gamma}}$ (randomly sampled) of GeSn alloys. The
ensemble-averaged band gaps $\langle E_g^{\Gamma}\rangle$ are obtained by randomly choosing
50 configurations within the canonical ensemble obtained from MC/DFT trajectory and averaging their
respective band gaps, thus reflecting the proper statistical weights
of the configurations and by default, taking into account the
identified SRO. The validity and effectiveness of
this procedure are further confirmed by a convergence test involving 500 configurations (See Supporting
Information S3 for more details). In contrast, the simply-averaged
band gaps $\overline{E_g^{\Gamma}}$ are obtained by
randomly generating 50 alloy configurations, from
which the band gaps are calculated and averaged. This procedure mimics
a completely random solid solution, and indeed reflects the essence
underlying the commonly employed theoretical approaches for modeling
random alloy, including VCA, CPA and SQS. 

\begin{figure}
\includegraphics[width=\linewidth]{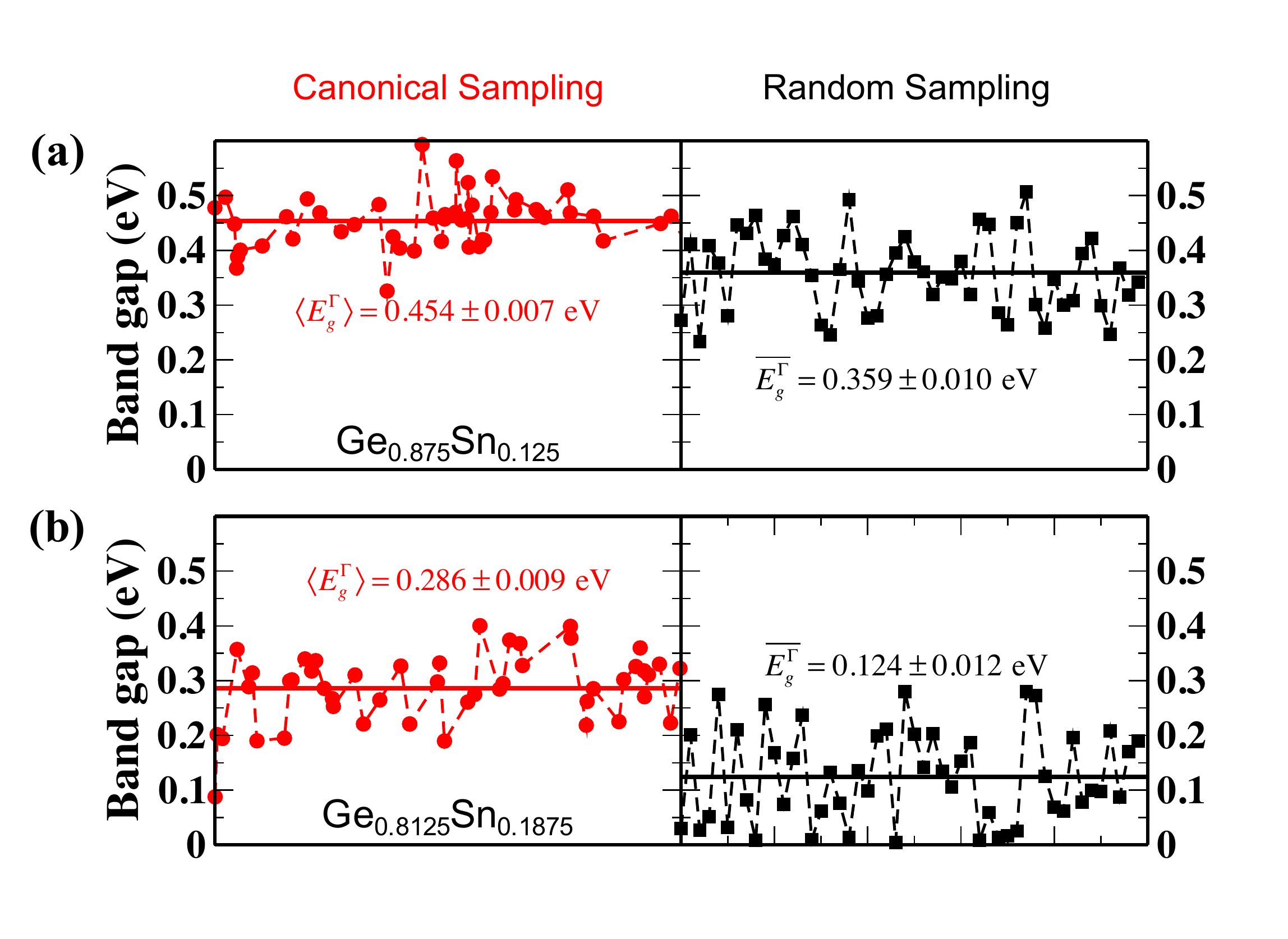}
\caption{\label{BGcomp} Effect of short-range order on the
  predicted direct (non-SOC) band gaps of GeSn alloy. The ensemble-averaged
  band gaps $\langle E_g^{\Gamma}\rangle$, obtained through canonical
  sampling (left column) are significantly higher than the
  simply-averaged band gaps $\overline{E_g^{\Gamma}}$, obtained through random
  sampling (right column), for (a) Ge$_{0.875}$Sn$_{0.125}$ and (b)
  Ge$_{0.8125}$Sn$_{0.1875}$. The horizontal lines represent the mean
  gap values. }
\end{figure}

Fig. \ref{BGcomp}(a) shows the comparison of $\langle E_g^{\Gamma}\rangle$ and
$\overline{E_g^{\Gamma}}$ for Ge$_{0.875}$Sn$_{0.125}$. It is evident
that the ensemble-averaged band gap $\langle E_g^{\Gamma}\rangle$ is
significantly higher than $\overline{E_g^{\Gamma}}$, by about 100 meV. In addition, the variation of
$E_g^{\Gamma}$, as reflected in the error bar of the mean, is also found
smaller in $\langle E_g^{\Gamma}\rangle$ than in
$\overline{E_g^{\Gamma}}$. Such differences clearly demonstrate the
quality of alloy structural models, particularly, whether the SRO is
taken into account, significantly affects the
predicted band gaps of GeSn alloys. 

To understand the fundamental origin for this difference, we
investigate its correlation with alloys' structures. To this end, it is worth pointing out
that a recent study \cite{Polak:2017fh} clearly demonstrated the
strong influence of Sn configurations on alloy's electronic band
gaps. In particular, the study showed the band gap exhibits the
maximum when Sn atoms are well separated (without nearest Sn-Sn
neighbors), then decreases monotonically as the number of Sn-Sn bonds
increases, and reaches the minimum when Sn atoms form a
cluster. Motivated by this observation, we examine the general relation
between band gaps and alloy structures. As shown in
Supporting Information Fig. S4, there indeed exists a general,
albeit not perfect, correlation between the number of
Sn-Sn bonds and the calculated band gaps: A higher number
Sn-Sn bonds overall tends to yield a lower band gap. Since an alloy configuration with a higher number of Sn-Sn bonds is
associated with a higher total energy (see Fig. \ref{dist}(b)),
such structure carries a lower statistical
weight thus contributes less to the canonical average. In other words,
the SRO plays an important role by limiting the
occurrence of those low-gap configurations in the 
ensemble. This explains why an ensemble-averaged band gap that
includes SRO leads to a higher band gap than
that inferred from the random solid solution model.  

In light of this understanding, and considering that the degree of
SRO in Ge-rich GeSn alloy grows
with Sn content, as shown in Fig. \ref{CNcomp}(a), one would expect the
correction in the predicted band gaps becomes even more significant for high-Sn
content alloy. This is because as Sn content increases, there will be
a higher probability of forming Sn-Sn nearest neighbors, or even
small Sn clusters, if atoms are randomly distributed within the
lattice. To confirm this conjecture, we compare the ensemble-averaged and
simply-averaged direct band gap for GeSn alloy
containing 18.75 at. \% Sn content. As shown by Fig. \ref{BGcomp}(b), the
difference between $\langle E_g^{\Gamma}\rangle$ and
$\overline{E_g^{\Gamma}}$ for Ge$_{0.8125}$Sn$_{0.1875}$ is indeed
found to increase to about 160 meV. 

\subsubsection{Comparison with experiments}

\begin{figure}
\includegraphics[width=0.75\linewidth]{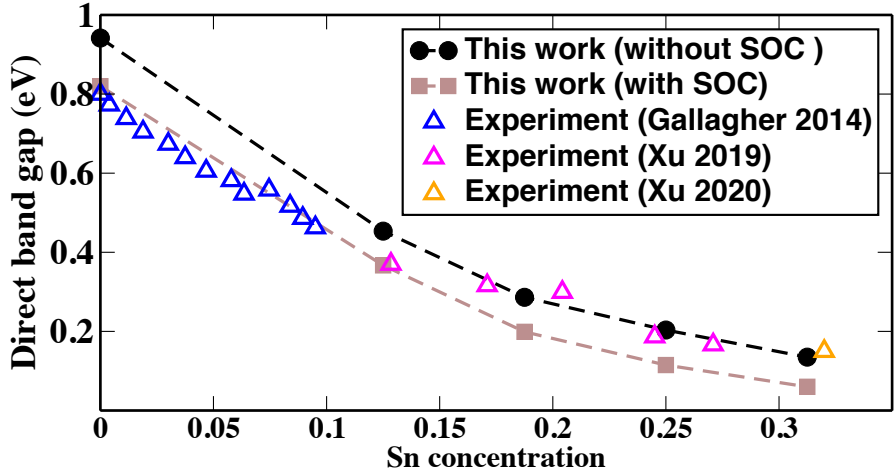}
\caption{\label{gapmap} Composition dependence of the direct band gap
  $E_g^{\Gamma}$ in GeSn alloy. The experimentally measured band gaps
  are from Gallagher 2014 \cite{Gallagher:2014jj}, Xu 2019
  \cite{Xu:2019ds}, and Xu 2019 \cite{Xu:2019bo}. }
\end{figure}

To gain a comprehensive understanding on the impact of SRO on
the quality of prediction for direct band gap in GeSn alloy, we carry out extensive sampling
and calculation to obtain the ensemble-averaged
direct band gap as a function of Sn composition. As shown in
Fig. \ref{gapmap}, our predicted direct band gaps show excellent
agreement with experiments \cite{Gallagher:2014jj,Xu:2019ds,Xu:2019bo}
within the entire composition range that has been visited by
experiments so far, with
the most salient improvements being for those high-Sn
content alloys with a Sn content beyond 20 at. \%. It is noted that
previous studies \cite{Eckhardt:2014gz,Polak:2017fh}, using the same level of DFT
  calculation, {\em i.e.}, mBJ and SOC, but assuming random solid solution, predicted GeSn alloy already
becomes a metal at around $25\sim28$ at. \% . Since high-Sn content alloys (with a Sn content beyond 20 at. \%) are of particular interest
for the proposed mid-infrared applications \cite{Li:2018ka}, our
results clearly show that constructing reliable structural models that
account for SRO is crucial for predicting and
understanding the electronic structures of those alloys. 
    
\section{Discussion}

Our results unravel a subtle but important difference between a
homogeneous GeSn alloy with a SRO where Sn atoms tend to avoid each
other as their immediate neighbors, and a truly random alloy where a lattice
distribution of atoms is completely randomized. Since both types of
distributions lead to a well
dispersion of Sn atoms in lattice, the absence of Sn segregation
cannot be used to infer either scenario. In fact the identified SRO
through Sn-Sn repulsion is expected to facilitate a better Sn dispersion, particularly for
high-Sn content GeSn alloy. To this end, we note that in addition to the demonstrated corrections to electronic structures,
our finding of SRO also has important implications on other
properties of GeSn alloys. One such implication is related to
understanding the
stability of GeSn alloy, as a
key challenge in achieving mid-infrared application of GeSn is
to ensure alloy is stable enough while incorporating sufficient Sn
content (beyond
20 at.\%) that is required for covering mid-infrared wavelength. Since
these compositions are significantly beyond the equilibrium
solubility of Sn in Ge ($\sim$ 1 at.\%) at room temperature, a major
expected issue is the segregation of Sn. Interestingly, despite of this
concern, high-Sn content GeSn alloys have been recently synthesized through
chemical-vapor deposition \cite{Dou:2018ik,Xu:2019ds,Xu:2019bo} and
molecular beam epitaxy \cite{Imbrenda:2018fw}. Although
strain gradient was found to play an important role in incorporating
Sn \cite{Dou:2018ik}, we conceive that an
additional main factor contributing to stability could well be related
to the identified lower-than-ideal Sn-Sn coordination number, because
a depletion of Sn atoms through a replacement by Ge
  atoms from their nearest neighbors could greatly reduce the
possibility of local gathering of Sn atoms, which is a prerequisite of
Sn segregation.

\section{Conclusion}

In summary, by combining Metropolis Monte Carlo sampling and
large-scale density
functional theory calculations, we reveal the existence of a
prominent, non-truly random solution behavior in GeSn alloy within the entire
composition range, in contrast to the commonly adopted model of random
solution. The calculated solute-solute
coordination numbers are found substantially lower than what is
assumed by the random solid solution model, demonstrating a partial, short-range order involving
solute atoms in GeSn.  The identified SRO is consistent with the EXAFS measurement
showing a lack of Sn-Sn nearest neighbors in GeSn alloys. When this short-range order is included in
modeling through ensemble average, we show the predicted band gaps of
GeSn alloys can be significantly improved through an excellent
agreement with recent experimental measurements, particularly for
high-Sn content GeSn alloys. The identified partial ordering is also
expected to play a vital role in understanding the optoelectronic properties and
stability of group IV alloys.  

\section{Supporting Information}

S1: Estimate of the statistical uncertainty in solute-solute
coordination number; S2: Effect of spin-orbit coupling on the
predicted direct band gap $E_g^{\Gamma}$; S3:
Convergence of sampling in computing average band gap $E_g^{\Gamma}$;
S4: Correlation between $E_g^{\Gamma}$ and number of Sn-Sn bonds.

\section*{Acknowledgment}

The authors thank Prof. Shui-Qing (Fisher) Yu and Dr. Enshi Xu for helpful discussions. This material is based upon work supported by the Air
Force Office of Scientific Research under award number
FA9550-19-1-0341. The authors acknowledge Department of Defense High Performance Computing Modernization Program for computing support.


\begin{mcitethebibliography}{55}
\providecommand*\natexlab[1]{#1}
\providecommand*\mciteSetBstSublistMode[1]{}
\providecommand*\mciteSetBstMaxWidthForm[2]{}
\providecommand*\mciteBstWouldAddEndPuncttrue
  {\def\EndOfBibitem{\unskip.}}
\providecommand*\mciteBstWouldAddEndPunctfalse
  {\let\EndOfBibitem\relax}
\providecommand*\mciteSetBstMidEndSepPunct[3]{}
\providecommand*\mciteSetBstSublistLabelBeginEnd[3]{}
\providecommand*\EndOfBibitem{}
\mciteSetBstSublistMode{f}
\mciteSetBstMaxWidthForm{subitem}{(\alph{mcitesubitemcount})}
\mciteSetBstSublistLabelBeginEnd
  {\mcitemaxwidthsubitemform\space}
  {\relax}
  {\relax}

\bibitem[Soref and Perry(1991)Soref, and Perry]{Soref:1991ce}
Soref,~R.~A.; Perry,~C.~H. Predicted Band Gap of the New Semiconductor
  {SiGeSn}. \emph{Journal of Applied Physics} \textbf{1991}, \emph{69},
  539--541\relax
\mciteBstWouldAddEndPuncttrue
\mciteSetBstMidEndSepPunct{\mcitedefaultmidpunct}
{\mcitedefaultendpunct}{\mcitedefaultseppunct}\relax
\EndOfBibitem
\bibitem[Gencarelli \latin{et~al.}(2013)Gencarelli, Vincent, Demeulemeester,
  Vantomme, Moussa, Franquet, Kumar, Bender, Meersschaut, Vandervorst, Loo,
  Caymax, Temst, and Heyns]{Gencarelli:2013go}
Gencarelli,~F.; Vincent,~B.; Demeulemeester,~J.; Vantomme,~A.; Moussa,~A.;
  Franquet,~A.; Kumar,~A.; Bender,~H.; Meersschaut,~J.; Vandervorst,~W.;
  Loo,~R.; Caymax,~M.; Temst,~K.; Heyns,~M. Crystalline {Properties} and
  {Strain} {Relaxation} {Mechanism} of {CVD} {Grown} {GeSn}. \emph{ECS J. Solid
  State Sci. Technol.} \textbf{2013}, \emph{2}, P134\relax
\mciteBstWouldAddEndPuncttrue
\mciteSetBstMidEndSepPunct{\mcitedefaultmidpunct}
{\mcitedefaultendpunct}{\mcitedefaultseppunct}\relax
\EndOfBibitem
\bibitem[Wirths \latin{et~al.}(2016)Wirths, Buca, and Mantl]{Wirths:2016ku}
Wirths,~S.; Buca,~D.; Mantl,~S. Si–{Ge}–{Sn} Alloys: {From} Growth to
  Applications. \emph{Progress in Crystal Growth and Characterization of
  Materials} \textbf{2016}, \emph{62}, 1--39\relax
\mciteBstWouldAddEndPuncttrue
\mciteSetBstMidEndSepPunct{\mcitedefaultmidpunct}
{\mcitedefaultendpunct}{\mcitedefaultseppunct}\relax
\EndOfBibitem
\bibitem[Ghetmiri \latin{et~al.}(2014)Ghetmiri, Du, Margetis, Mosleh, Cousar,
  Conley, Domulevicz, Nazzal, Sun, Soref, Tolle, Li, Naseem, and
  Yu]{Ghetmiri:2014hh}
Ghetmiri,~S.~A.; Du,~W.; Margetis,~J.; Mosleh,~A.; Cousar,~L.; Conley,~B.~R.;
  Domulevicz,~L.; Nazzal,~A.; Sun,~G.; Soref,~R.~A.; Tolle,~J.; Li,~B.;
  Naseem,~H.~A.; Yu,~S.-Q. Direct-Bandgap {GeSn} Grown on Silicon with 2230 nm
  Photoluminescence. \emph{Appl. Phys. Lett.} \textbf{2014}, \emph{105},
  151109\relax
\mciteBstWouldAddEndPuncttrue
\mciteSetBstMidEndSepPunct{\mcitedefaultmidpunct}
{\mcitedefaultendpunct}{\mcitedefaultseppunct}\relax
\EndOfBibitem
\bibitem[Wirths \latin{et~al.}(2015)Wirths, Geiger, von~den Driesch, Mussler,
  Stoica, Mantl, Ikonic, Luysberg, Chiussi, Hartmann, Sigg, Faist, Buca, and
  Grützmacher]{Wirths:2015dh}
Wirths,~S.; Geiger,~R.; von~den Driesch,~N.; Mussler,~G.; Stoica,~T.;
  Mantl,~S.; Ikonic,~Z.; Luysberg,~M.; Chiussi,~S.; Hartmann,~J.~M.; Sigg,~H.;
  Faist,~J.; Buca,~D.; Grützmacher,~D. Lasing in Direct-Bandgap {GeSn} Alloy
  Grown on {Si}. \emph{Nature Photonics} \textbf{2015}, \emph{9}, 88--92\relax
\mciteBstWouldAddEndPuncttrue
\mciteSetBstMidEndSepPunct{\mcitedefaultmidpunct}
{\mcitedefaultendpunct}{\mcitedefaultseppunct}\relax
\EndOfBibitem
\bibitem[Stange \latin{et~al.}(2016)Stange, Wirths, Geiger, Schulte-Braucks,
  Marzban, von~den Driesch, Mussler, Zabel, Stoica, Hartmann, Mantl, Ikonic,
  Grützmacher, Sigg, Witzens, and Buca]{Stange:2016ip}
Stange,~D.; Wirths,~S.; Geiger,~R.; Schulte-Braucks,~C.; Marzban,~B.; von~den
  Driesch,~N.; Mussler,~G.; Zabel,~T.; Stoica,~T.; Hartmann,~J.-M.; Mantl,~S.;
  Ikonic,~Z.; Grützmacher,~D.; Sigg,~H.; Witzens,~J.; Buca,~D. Optically
  {Pumped} {GeSn} {Microdisk} {Lasers} on {Si}. \emph{ACS Photonics}
  \textbf{2016}, \emph{3}, 1279--1285\relax
\mciteBstWouldAddEndPuncttrue
\mciteSetBstMidEndSepPunct{\mcitedefaultmidpunct}
{\mcitedefaultendpunct}{\mcitedefaultseppunct}\relax
\EndOfBibitem
\bibitem[Reboud \latin{et~al.}(2017)Reboud, Gassenq, Pauc, Aubin, Milord, Thai,
  Bertrand, Guilloy, Rouchon, Rothman, Zabel, Armand~Pilon, Sigg, Chelnokov,
  Hartmann, and Calvo]{Reboud:2017cr}
Reboud,~V.; Gassenq,~A.; Pauc,~N.; Aubin,~J.; Milord,~L.; Thai,~Q.~M.;
  Bertrand,~M.; Guilloy,~K.; Rouchon,~D.; Rothman,~J.; Zabel,~T.;
  Armand~Pilon,~F.; Sigg,~H.; Chelnokov,~A.; Hartmann,~J.~M.; Calvo,~V.
  Optically Pumped {GeSn} Micro-Disks with 16\% {Sn} Lasing at 3.1 $\mu$m up to
  180 {K}. \emph{Appl. Phys. Lett.} \textbf{2017}, \emph{111}, 092101\relax
\mciteBstWouldAddEndPuncttrue
\mciteSetBstMidEndSepPunct{\mcitedefaultmidpunct}
{\mcitedefaultendpunct}{\mcitedefaultseppunct}\relax
\EndOfBibitem
\bibitem[Al-Kabi \latin{et~al.}(2016)Al-Kabi, Ghetmiri, Margetis, Pham, Zhou,
  Dou, Collier, Quinde, Du, Mosleh, Liu, Sun, Soref, Tolle, Li, Mortazavi,
  Naseem, and Yu]{AlKabi:2016de}
Al-Kabi,~S.; Ghetmiri,~S.~A.; Margetis,~J.; Pham,~T.; Zhou,~Y.; Dou,~W.;
  Collier,~B.; Quinde,~R.; Du,~W.; Mosleh,~A.; Liu,~J.; Sun,~G.; Soref,~R.~A.;
  Tolle,~J.; Li,~B.; Mortazavi,~M.; Naseem,~H.~A.; Yu,~S.-Q. An Optically
  Pumped 2.5 $\mu$m {GeSn} Laser on {Si} Operating at 110 {K}. \emph{Appl.
  Phys. Lett.} \textbf{2016}, \emph{109}, 171105\relax
\mciteBstWouldAddEndPuncttrue
\mciteSetBstMidEndSepPunct{\mcitedefaultmidpunct}
{\mcitedefaultendpunct}{\mcitedefaultseppunct}\relax
\EndOfBibitem
\bibitem[Margetis \latin{et~al.}(2018)Margetis, Al-Kabi, Du, Dou, Zhou, Pham,
  Grant, Ghetmiri, Mosleh, Li, Liu, Sun, Soref, Tolle, Mortazavi, and
  Yu]{Margetis:2017fs}
Margetis,~J.; Al-Kabi,~S.; Du,~W.; Dou,~W.; Zhou,~Y.; Pham,~T.; Grant,~P.;
  Ghetmiri,~S.; Mosleh,~A.; Li,~B.; Liu,~J.; Sun,~G.; Soref,~R.; Tolle,~J.;
  Mortazavi,~M.; Yu,~S.-Q. Si-{Based} {GeSn} {Lasers} with {Wavelength}
  {Coverage} of 2–3 $\mu$m and {Operating} {Temperatures} up to 180 {K}.
  \emph{ACS Photonics} \textbf{2018}, \emph{5}, 827--833\relax
\mciteBstWouldAddEndPuncttrue
\mciteSetBstMidEndSepPunct{\mcitedefaultmidpunct}
{\mcitedefaultendpunct}{\mcitedefaultseppunct}\relax
\EndOfBibitem
\bibitem[Dou \latin{et~al.}(2018)Dou, Zhou, Margetis, Ghetmiri, Al-Kabi, Du,
  Liu, Sun, Soref, Tolle, Li, Mortazavi, and Yu]{Li:2018ka}
Dou,~W.; Zhou,~Y.; Margetis,~J.; Ghetmiri,~S.~A.; Al-Kabi,~S.; Du,~W.; Liu,~J.;
  Sun,~G.; Soref,~R.~A.; Tolle,~J.; Li,~B.; Mortazavi,~M.; Yu,~S.-Q. Optically
  Pumped Lasing at 3 $\mu$m from Compositionally Graded {GeSn} with Tin up to
  22.3\%. \emph{Opt. Lett.} \textbf{2018}, \emph{43}, 4558--4561\relax
\mciteBstWouldAddEndPuncttrue
\mciteSetBstMidEndSepPunct{\mcitedefaultmidpunct}
{\mcitedefaultendpunct}{\mcitedefaultseppunct}\relax
\EndOfBibitem
\bibitem[Soref(2010)]{Soref:2010iw}
Soref,~R. Mid-Infrared Photonics in Silicon and Germanium. \emph{Nature
  Photonics} \textbf{2010}, \emph{4}, 495--497\relax
\mciteBstWouldAddEndPuncttrue
\mciteSetBstMidEndSepPunct{\mcitedefaultmidpunct}
{\mcitedefaultendpunct}{\mcitedefaultseppunct}\relax
\EndOfBibitem
\bibitem[Kumar \latin{et~al.}(2013)Kumar, Komalan, Lenka, Kambham, Gilbert,
  Gencarelli, Vincent, and Vandervorst]{Kumar:2013ih}
Kumar,~A.; Komalan,~M.~P.; Lenka,~H.; Kambham,~A.~K.; Gilbert,~M.;
  Gencarelli,~F.; Vincent,~B.; Vandervorst,~W. {Atomic insight into
  Ge$_{\text{1-x}}$Sn$_{\text{x}}$ using atom probe tomography}.
  \emph{Ultramicroscopy} \textbf{2013}, \emph{132}, 171--178\relax
\mciteBstWouldAddEndPuncttrue
\mciteSetBstMidEndSepPunct{\mcitedefaultmidpunct}
{\mcitedefaultendpunct}{\mcitedefaultseppunct}\relax
\EndOfBibitem
\bibitem[Kumar \latin{et~al.}(2015)Kumar, Demeulemeester, Bogdanowicz, Bran,
  Melkonyan, Fleischmann, Gencarelli, Shimura, Wang, Loo, and
  Vandervorst]{Kumar:2015ha}
Kumar,~A.; Demeulemeester,~J.; Bogdanowicz,~J.; Bran,~J.; Melkonyan,~D.;
  Fleischmann,~C.; Gencarelli,~F.; Shimura,~Y.; Wang,~W.; Loo,~R.;
  Vandervorst,~W. {On the interplay between relaxation, defect formation, and
  atomic Sn distribution in Ge$_{(1-x)}$Sn$_{(x)}$ unraveled with atom probe
  tomography}. \emph{J Appl Phys} \textbf{2015}, \emph{118}, 025302\relax
\mciteBstWouldAddEndPuncttrue
\mciteSetBstMidEndSepPunct{\mcitedefaultmidpunct}
{\mcitedefaultendpunct}{\mcitedefaultseppunct}\relax
\EndOfBibitem
\bibitem[Assali \latin{et~al.}(2017)Assali, Dijkstra, Li, Koelling, Verheijen,
  Gagliano, von~den Driesch, Buca, Koenraad, Haverkort, and
  Bakkers]{Assali:2017et}
Assali,~S.; Dijkstra,~A.; Li,~A.; Koelling,~S.; Verheijen,~M.~A.; Gagliano,~L.;
  von~den Driesch,~N.; Buca,~D.; Koenraad,~P.~M.; Haverkort,~J. E.~M.;
  Bakkers,~E. P. A.~M. {Growth and Optical Properties of Direct Band Gap
  Ge/Ge$_{0.87}$Sn$_{0.13}$ Core/Shell Nanowire Arrays}. \emph{Nano Letters}
  \textbf{2017}, \emph{17}, 1538--1544\relax
\mciteBstWouldAddEndPuncttrue
\mciteSetBstMidEndSepPunct{\mcitedefaultmidpunct}
{\mcitedefaultendpunct}{\mcitedefaultseppunct}\relax
\EndOfBibitem
\bibitem[Assali \latin{et~al.}(2018)Assali, Nicolas, Mukherjee, Dijkstra, and
  Moutanabbir]{Assali:2018db}
Assali,~S.; Nicolas,~J.; Mukherjee,~S.; Dijkstra,~A.; Moutanabbir,~O.
  {Atomically Uniform {Sn}-Rich {GeSn} Semiconductors with 3.0--3.5 $\mu$m
  Room-Temperature Optical Emission}. \emph{Appl. Phys. Lett.} \textbf{2018},
  \emph{112}, 251903\relax
\mciteBstWouldAddEndPuncttrue
\mciteSetBstMidEndSepPunct{\mcitedefaultmidpunct}
{\mcitedefaultendpunct}{\mcitedefaultseppunct}\relax
\EndOfBibitem
\bibitem[Xu \latin{et~al.}(2019)Xu, Ringwala, Wang, Liu, Poweleit, Chang,
  Zhuang, Menéndez, and Kouvetakis]{Xu:2019bo}
Xu,~C.; Ringwala,~D.; Wang,~D.; Liu,~L.; Poweleit,~C.~D.; Chang,~S. L.~Y.;
  Zhuang,~H.~L.; Menéndez,~J.; Kouvetakis,~J. Synthesis and {Fundamental}
  {Studies} of {Si}-{Compatible} ({Si}){GeSn} and {GeSn} {Mid}-{IR} {Systems}
  with {Ultrahigh} {Sn} {Contents}. \emph{Chem. Mater.} \textbf{2019},
  \emph{31}, 9831--9842\relax
\mciteBstWouldAddEndPuncttrue
\mciteSetBstMidEndSepPunct{\mcitedefaultmidpunct}
{\mcitedefaultendpunct}{\mcitedefaultseppunct}\relax
\EndOfBibitem
\bibitem[Doherty \latin{et~al.}(2020)Doherty, Biswas, Galluccio, Broderick,
  Garcia-Gil, Duffy, O’Reilly, and Holmes]{Doherty:2020id}
Doherty,~J.; Biswas,~S.; Galluccio,~E.; Broderick,~C.~A.; Garcia-Gil,~A.;
  Duffy,~R.; O’Reilly,~E.~P.; Holmes,~J.~D. Progress on {Germanium}–{Tin}
  {Nanoscale} {Alloys}. \emph{Chem. Mater.} \textbf{2020}, \emph{32},
  4383--4408\relax
\mciteBstWouldAddEndPuncttrue
\mciteSetBstMidEndSepPunct{\mcitedefaultmidpunct}
{\mcitedefaultendpunct}{\mcitedefaultseppunct}\relax
\EndOfBibitem
\bibitem[Shen \latin{et~al.}(1997)Shen, Zi, Xie, and Jiang]{Shen:1997fc}
Shen,~J.; Zi,~J.; Xie,~X.; Jiang,~P. Ab Initio Calculation of the Structure of
  the Random Alloys Sn$_x$Ge$_{1-x}$. \emph{Phys. Rev. B} \textbf{1997},
  \emph{56}, 12084--12087\relax
\mciteBstWouldAddEndPuncttrue
\mciteSetBstMidEndSepPunct{\mcitedefaultmidpunct}
{\mcitedefaultendpunct}{\mcitedefaultseppunct}\relax
\EndOfBibitem
\bibitem[Pandey \latin{et~al.}(1999)Pandey, Rérat, and Causà]{Pandey:1999ii}
Pandey,~R.; Rérat,~M.; Causà,~M. First-Principles Study of Stability, Band
  Structure, and Optical Properties of the Ordered Ge$_{0.50}$Sn$_{0.50}$
  Alloy. \emph{Appl. Phys. Lett.} \textbf{1999}, \emph{75}, 4127--4129\relax
\mciteBstWouldAddEndPuncttrue
\mciteSetBstMidEndSepPunct{\mcitedefaultmidpunct}
{\mcitedefaultendpunct}{\mcitedefaultseppunct}\relax
\EndOfBibitem
\bibitem[Moontragoon \latin{et~al.}(2007)Moontragoon, Ikonić, and
  Harrison]{Moontragoon:2007ho}
Moontragoon,~P.; Ikonić,~Z.; Harrison,~P. Band Structure Calculations of
  {Si}–{Ge}–{Sn} Alloys: Achieving Direct Band Gap Materials.
  \emph{Semicond. Sci. Technol.} \textbf{2007}, \emph{22}, 742--748\relax
\mciteBstWouldAddEndPuncttrue
\mciteSetBstMidEndSepPunct{\mcitedefaultmidpunct}
{\mcitedefaultendpunct}{\mcitedefaultseppunct}\relax
\EndOfBibitem
\bibitem[Yin \latin{et~al.}(2008)Yin, Gong, and Wei]{Yin:2008cf}
Yin,~W.-J.; Gong,~X.-G.; Wei,~S.-H. Origin of the Unusually Large Band-Gap
  Bowing and the Breakdown of the Band-Edge Distribution Rule in the
  Sn$_x$Ge$_{1-x}$ Alloys. \emph{Phys. Rev. B} \textbf{2008}, \emph{78},
  161203\relax
\mciteBstWouldAddEndPuncttrue
\mciteSetBstMidEndSepPunct{\mcitedefaultmidpunct}
{\mcitedefaultendpunct}{\mcitedefaultseppunct}\relax
\EndOfBibitem
\bibitem[Beeler \latin{et~al.}(2011)Beeler, Roucka, Chizmeshya, Kouvetakis, and
  Menéndez]{Beeler:2011be}
Beeler,~R.; Roucka,~R.; Chizmeshya,~A. V.~G.; Kouvetakis,~J.; Menéndez,~J.
  Nonlinear Structure-Composition Relationships in the
  Ge$_{1-y}$Sn$_y$/{Si}(100) (y {\textless} 0.15) System. \emph{Phys. Rev. B}
  \textbf{2011}, \emph{84}, 035204\relax
\mciteBstWouldAddEndPuncttrue
\mciteSetBstMidEndSepPunct{\mcitedefaultmidpunct}
{\mcitedefaultendpunct}{\mcitedefaultseppunct}\relax
\EndOfBibitem
\bibitem[Lu~Low \latin{et~al.}(2012)Lu~Low, Yang, Han, Fan, and
  Yeo]{Low:2012bh}
Lu~Low,~K.; Yang,~Y.; Han,~G.; Fan,~W.; Yeo,~Y.-C. Electronic Band Structure
  and Effective Mass Parameters of Ge$_{1-x}$Sn$_x$ Alloys. \emph{Journal of
  Applied Physics} \textbf{2012}, \emph{112}, 103715\relax
\mciteBstWouldAddEndPuncttrue
\mciteSetBstMidEndSepPunct{\mcitedefaultmidpunct}
{\mcitedefaultendpunct}{\mcitedefaultseppunct}\relax
\EndOfBibitem
\bibitem[Lee \latin{et~al.}(2013)Lee, Liu, Hong, Chou, Hong, and
  Siao]{Lee:2013ib}
Lee,~M.-H.; Liu,~P.-L.; Hong,~Y.-A.; Chou,~Y.-T.; Hong,~J.-Y.; Siao,~Y.-J.
  Electronic Band Structures of Ge$_{1-x}$Sn$_x$ Semiconductors: {A}
  First-Principles Density Functional Theory Study. \emph{Journal of Applied
  Physics} \textbf{2013}, \emph{113}, 063517\relax
\mciteBstWouldAddEndPuncttrue
\mciteSetBstMidEndSepPunct{\mcitedefaultmidpunct}
{\mcitedefaultendpunct}{\mcitedefaultseppunct}\relax
\EndOfBibitem
\bibitem[Gupta \latin{et~al.}(2013)Gupta, Magyari-Köpe, Nishi, and
  Saraswat]{Gupta:2013kp}
Gupta,~S.; Magyari-Köpe,~B.; Nishi,~Y.; Saraswat,~K.~C. Achieving Direct Band
  Gap in Germanium Through Integration of {Sn} Alloying and External Strain.
  \emph{Journal of Applied Physics} \textbf{2013}, \emph{113}, 073707\relax
\mciteBstWouldAddEndPuncttrue
\mciteSetBstMidEndSepPunct{\mcitedefaultmidpunct}
{\mcitedefaultendpunct}{\mcitedefaultseppunct}\relax
\EndOfBibitem
\bibitem[Eckhardt \latin{et~al.}(2014)Eckhardt, Hummer, and
  Kresse]{Eckhardt:2014gz}
Eckhardt,~C.; Hummer,~K.; Kresse,~G. Indirect-To-Direct Gap Transition in
  Strained and Unstrained Sn$_x$Ge$_{1-x}$ Alloys. \emph{Phys. Rev. B}
  \textbf{2014}, \emph{89}, 165201\relax
\mciteBstWouldAddEndPuncttrue
\mciteSetBstMidEndSepPunct{\mcitedefaultmidpunct}
{\mcitedefaultendpunct}{\mcitedefaultseppunct}\relax
\EndOfBibitem
\bibitem[Freitas \latin{et~al.}(2016)Freitas, Furthmüller, Bechstedt, Marques,
  and Teles]{Freitas:2016kc}
Freitas,~F.~L.; Furthmüller,~J.; Bechstedt,~F.; Marques,~M.; Teles,~L.~K.
  Influence of the Composition Fluctuations and Decomposition on the Tunable
  Direct Gap and Oscillator Strength of Ge$_{1-x}$Sn$_x$ Alloys. \emph{Appl.
  Phys. Lett.} \textbf{2016}, \emph{108}, 092101\relax
\mciteBstWouldAddEndPuncttrue
\mciteSetBstMidEndSepPunct{\mcitedefaultmidpunct}
{\mcitedefaultendpunct}{\mcitedefaultseppunct}\relax
\EndOfBibitem
\bibitem[Polak \latin{et~al.}(2017)Polak, Scharoch, and
  Kudrawiec]{Polak:2017fh}
Polak,~M.~P.; Scharoch,~P.; Kudrawiec,~R. The Electronic Band Structure of
  {Ge1}-{xSnxin} the Full Composition Range: Indirect, Direct, and Inverted
  Gaps Regimes, Band Offsets, and the {Burstein}–{Moss} Effect. \emph{J.
  Phys. D: Appl. Phys.} \textbf{2017}, \emph{50}, 195103\relax
\mciteBstWouldAddEndPuncttrue
\mciteSetBstMidEndSepPunct{\mcitedefaultmidpunct}
{\mcitedefaultendpunct}{\mcitedefaultseppunct}\relax
\EndOfBibitem
\bibitem[O'Halloran \latin{et~al.}(2019)O'Halloran, Broderick, Tanner, Schulz,
  and O’Reilly]{OHalloran:2019ik}
O'Halloran,~E.~J.; Broderick,~C.~A.; Tanner,~D. S.~P.; Schulz,~S.;
  O’Reilly,~E.~P. Comparison of First Principles and Semi-Empirical Models of
  the Structural and Electronic Properties of Ge$_{1-x}$Sn$_x$ Alloys.
  \emph{Opt Quant Electron} \textbf{2019}, \emph{51}, 314\relax
\mciteBstWouldAddEndPuncttrue
\mciteSetBstMidEndSepPunct{\mcitedefaultmidpunct}
{\mcitedefaultendpunct}{\mcitedefaultseppunct}\relax
\EndOfBibitem
\bibitem[Zunger \latin{et~al.}(1990)Zunger, Wei, Ferreira, and
  Bernard]{Zunger:1990tw}
Zunger,~A.; Wei,~S.-H.; Ferreira,~L.~G.; Bernard,~J.~E. Special Quasirandom
  Structures. \emph{Phys. Rev. Lett.} \textbf{1990}, \emph{65}, 353--356\relax
\mciteBstWouldAddEndPuncttrue
\mciteSetBstMidEndSepPunct{\mcitedefaultmidpunct}
{\mcitedefaultendpunct}{\mcitedefaultseppunct}\relax
\EndOfBibitem
\bibitem[Gencarelli \latin{et~al.}(2015)Gencarelli, Grandjean, Shimura,
  Vincent, Banerjee, Vantomme, Vandervorst, Loo, Heyns, and
  Temst]{Gencarelli:2015fl}
Gencarelli,~F.; Grandjean,~D.; Shimura,~Y.; Vincent,~B.; Banerjee,~D.;
  Vantomme,~A.; Vandervorst,~W.; Loo,~R.; Heyns,~M.; Temst,~K. Extended {X}-Ray
  Absorption Fine Structure Investigation of {Sn} Local Environment in Strained
  and Relaxed Epitaxial Ge$_{1-x}$Sn$_x$ Films. \emph{Journal of Applied
  Physics} \textbf{2015}, \emph{117}, 095702\relax
\mciteBstWouldAddEndPuncttrue
\mciteSetBstMidEndSepPunct{\mcitedefaultmidpunct}
{\mcitedefaultendpunct}{\mcitedefaultseppunct}\relax
\EndOfBibitem
\bibitem[Robouch \latin{et~al.}(2020)Robouch, Valeev, Kisiel, and
  Marcelli]{Robouch:2020ib}
Robouch,~B.~V.; Valeev,~R.~G.; Kisiel,~A.; Marcelli,~A. Atomic Distributions
  Observed in Group {IV}-{IV} Binary Tetrahedron Alloys: {A} Revised Analysis
  of {SiGe} and {GeSn} Compounds. \emph{Journal of Alloys and Compounds}
  \textbf{2020}, \emph{831}, 154743\relax
\mciteBstWouldAddEndPuncttrue
\mciteSetBstMidEndSepPunct{\mcitedefaultmidpunct}
{\mcitedefaultendpunct}{\mcitedefaultseppunct}\relax
\EndOfBibitem
\bibitem[Mukherjee \latin{et~al.}(2017)Mukherjee, Kodali, Isheim, Wirths,
  Hartmann, Buca, Seidman, and Moutanabbir]{Mukherjee:2017is}
Mukherjee,~S.; Kodali,~N.; Isheim,~D.; Wirths,~S.; Hartmann,~J.~M.; Buca,~D.;
  Seidman,~D.~N.; Moutanabbir,~O. Short-Range Atomic Ordering in Nonequilibrium
  Silicon-Germanium-Tin Semiconductors. \emph{Phys. Rev. B} \textbf{2017},
  \emph{95}, 161402\relax
\mciteBstWouldAddEndPuncttrue
\mciteSetBstMidEndSepPunct{\mcitedefaultmidpunct}
{\mcitedefaultendpunct}{\mcitedefaultseppunct}\relax
\EndOfBibitem
\bibitem[Xu \latin{et~al.}(2019)Xu, Wallace, Ringwala, Chang, Poweleit,
  Kouvetakis, and Menéndez]{Xu:2019ds}
Xu,~C.; Wallace,~P.~M.; Ringwala,~D.~A.; Chang,~S. L.~Y.; Poweleit,~C.~D.;
  Kouvetakis,~J.; Menéndez,~J. Mid-Infrared (3–8 $\mu$m) Ge$_{1-y}$Sn$_y$
  Alloys (0.15 {\textless} y {\textless} 0.30): {Synthesis}, Structural, and
  Optical Properties. \emph{Appl. Phys. Lett.} \textbf{2019}, \emph{114},
  212104\relax
\mciteBstWouldAddEndPuncttrue
\mciteSetBstMidEndSepPunct{\mcitedefaultmidpunct}
{\mcitedefaultendpunct}{\mcitedefaultseppunct}\relax
\EndOfBibitem
\bibitem[Dou \latin{et~al.}(2018)Dou, Benamara, Mosleh, Margetis, Grant, Zhou,
  Al-Kabi, Du, Tolle, Li, Mortazavi, and Yu]{Dou:2018ik}
Dou,~W.; Benamara,~M.; Mosleh,~A.; Margetis,~J.; Grant,~P.; Zhou,~Y.;
  Al-Kabi,~S.; Du,~W.; Tolle,~J.; Li,~B.; Mortazavi,~M.; Yu,~S.-Q.
  Investigation of {GeSn} {Strain} {Relaxation} and {Spontaneous} {Composition}
  {Gradient} for {Low}-{Defect} and {High}-{Sn} {Alloy} {Growth}.
  \emph{Scientific Reports} \textbf{2018}, \emph{8}, 5640\relax
\mciteBstWouldAddEndPuncttrue
\mciteSetBstMidEndSepPunct{\mcitedefaultmidpunct}
{\mcitedefaultendpunct}{\mcitedefaultseppunct}\relax
\EndOfBibitem
\bibitem[Imbrenda \latin{et~al.}(2018)Imbrenda, Hickey, Carrasco, Fernando,
  VanDerslice, Zollner, and Kolodzey]{Imbrenda:2018fw}
Imbrenda,~D.; Hickey,~R.; Carrasco,~R.~A.; Fernando,~N.~S.; VanDerslice,~J.;
  Zollner,~S.; Kolodzey,~J. Infrared Dielectric Response, Index of Refraction,
  and Absorption of Germanium-Tin Alloys with Tin Contents up to 27\% Deposited
  by Molecular Beam Epitaxy. \emph{Appl. Phys. Lett.} \textbf{2018},
  \emph{113}, 122104\relax
\mciteBstWouldAddEndPuncttrue
\mciteSetBstMidEndSepPunct{\mcitedefaultmidpunct}
{\mcitedefaultendpunct}{\mcitedefaultseppunct}\relax
\EndOfBibitem
\bibitem[Zhang \latin{et~al.}(2017)Zhang, Zhao, Jin, Xue, Velisa, Bei, Huang,
  Ko, Pagan, Neuefeind, Weber, and Zhang]{Zhang:2017bw}
Zhang,~F.; Zhao,~S.; Jin,~K.; Xue,~H.; Velisa,~G.; Bei,~H.; Huang,~R.; Ko,~J.;
  Pagan,~D.; Neuefeind,~J.; Weber,~W.; Zhang,~Y. Local {Structure} and
  {Short}-{Range} {Order} in a {NiCoCr} {Solid} {Solution} {Alloy}. \emph{Phys.
  Rev. Lett.} \textbf{2017}, \emph{118}, 205501\relax
\mciteBstWouldAddEndPuncttrue
\mciteSetBstMidEndSepPunct{\mcitedefaultmidpunct}
{\mcitedefaultendpunct}{\mcitedefaultseppunct}\relax
\EndOfBibitem
\bibitem[Ding \latin{et~al.}(2018)Ding, Yu, Asta, and Ritchie]{Ding:2018im}
Ding,~J.; Yu,~Q.; Asta,~M.; Ritchie,~R.~O. Tunable Stacking Fault Energies by
  Tailoring Local Chemical Order in {CrCoNi} Medium-Entropy Alloys. \emph{PNAS}
  \textbf{2018}, \emph{115}, 8919--8924\relax
\mciteBstWouldAddEndPuncttrue
\mciteSetBstMidEndSepPunct{\mcitedefaultmidpunct}
{\mcitedefaultendpunct}{\mcitedefaultseppunct}\relax
\EndOfBibitem
\bibitem[Zhang \latin{et~al.}(2020)Zhang, Zhao, Ding, Chong, Jia, Ophus, Asta,
  Ritchie, and Minor]{Zhang:2020kr}
Zhang,~R.; Zhao,~S.; Ding,~J.; Chong,~Y.; Jia,~T.; Ophus,~C.; Asta,~M.;
  Ritchie,~R.~O.; Minor,~A.~M. Short-Range Order and Its Impact on the {CrCoNi}
  Medium-Entropy Alloy. \emph{Nature} \textbf{2020}, \emph{581}, 283--287\relax
\mciteBstWouldAddEndPuncttrue
\mciteSetBstMidEndSepPunct{\mcitedefaultmidpunct}
{\mcitedefaultendpunct}{\mcitedefaultseppunct}\relax
\EndOfBibitem
\bibitem[Metropolis \latin{et~al.}(1953)Metropolis, Rosenbluth, Rosenbluth,
  Teller, and Teller]{Metropolis:1953in}
Metropolis,~N.; Rosenbluth,~A.~W.; Rosenbluth,~M.~N.; Teller,~A.~H.; Teller,~E.
  Equation of {State} {Calculations} by {Fast} {Computing} {Machines}. \emph{J.
  Chem. Phys.} \textbf{1953}, \emph{21}, 1087--1092\relax
\mciteBstWouldAddEndPuncttrue
\mciteSetBstMidEndSepPunct{\mcitedefaultmidpunct}
{\mcitedefaultendpunct}{\mcitedefaultseppunct}\relax
\EndOfBibitem
\bibitem[Kresse and Hafner(1993)Kresse, and Hafner]{Kresse:1993ty}
Kresse,~G.; Hafner,~J. Ab Initio Molecular Dynamics for Liquid Metals.
  \emph{Phys. Rev. B} \textbf{1993}, \emph{47}, 558--561\relax
\mciteBstWouldAddEndPuncttrue
\mciteSetBstMidEndSepPunct{\mcitedefaultmidpunct}
{\mcitedefaultendpunct}{\mcitedefaultseppunct}\relax
\EndOfBibitem
\bibitem[Kresse and Joubert(1999)Kresse, and Joubert]{Kresse:1999tq}
Kresse,~G.; Joubert,~D. From Ultrasoft Pseudopotentials to the Projector
  Augmented-Wave Method. \emph{Phys. Rev. B} \textbf{1999}, \emph{59},
  1758--1775\relax
\mciteBstWouldAddEndPuncttrue
\mciteSetBstMidEndSepPunct{\mcitedefaultmidpunct}
{\mcitedefaultendpunct}{\mcitedefaultseppunct}\relax
\EndOfBibitem
\bibitem[Kresse and Furthmüller(1996)Kresse, and Furthmüller]{Kresse:1996kg}
Kresse,~G.; Furthmüller,~J. Efficiency of Ab-Initio Total Energy Calculations
  for Metals and Semiconductors Using a Plane-Wave Basis Set.
  \emph{Computational Materials Science} \textbf{1996}, \emph{6}, 15--50\relax
\mciteBstWouldAddEndPuncttrue
\mciteSetBstMidEndSepPunct{\mcitedefaultmidpunct}
{\mcitedefaultendpunct}{\mcitedefaultseppunct}\relax
\EndOfBibitem
\bibitem[Kresse and Furthmüller(1996)Kresse, and Furthmüller]{Kresse:1996vf}
Kresse,~G.; Furthmüller,~J. Efficient Iterative Schemes for Ab Initio
  Total-Energy Calculations Using a Plane-Wave Basis Set. \emph{Phys. Rev. B}
  \textbf{1996}, \emph{54}, 11169--11186\relax
\mciteBstWouldAddEndPuncttrue
\mciteSetBstMidEndSepPunct{\mcitedefaultmidpunct}
{\mcitedefaultendpunct}{\mcitedefaultseppunct}\relax
\EndOfBibitem
\bibitem[Ceperley and Alder(1980)Ceperley, and Alder]{Ceperley:1980gc}
Ceperley,~D.~M.; Alder,~B.~J. Ground {State} of the {Electron} {Gas} by a
  {Stochastic} {Method}. \emph{Phys. Rev. Lett.} \textbf{1980}, \emph{45},
  566--569\relax
\mciteBstWouldAddEndPuncttrue
\mciteSetBstMidEndSepPunct{\mcitedefaultmidpunct}
{\mcitedefaultendpunct}{\mcitedefaultseppunct}\relax
\EndOfBibitem
\bibitem[Haas \latin{et~al.}(2009)Haas, Tran, and Blaha]{Haas:2009be}
Haas,~P.; Tran,~F.; Blaha,~P. Calculation of the Lattice Constant of Solids
  with Semilocal Functionals. \emph{Phys. Rev. B} \textbf{2009}, \emph{79},
  085104\relax
\mciteBstWouldAddEndPuncttrue
\mciteSetBstMidEndSepPunct{\mcitedefaultmidpunct}
{\mcitedefaultendpunct}{\mcitedefaultseppunct}\relax
\EndOfBibitem
\bibitem[Perdew \latin{et~al.}(1996)Perdew, Burke, and
  Ernzerhof]{PERDEW:1996ug}
Perdew,~J.~P.; Burke,~K.; Ernzerhof,~M. Generalized {Gradient} {Approximation}
  {Made} {Simple}. \emph{Phys. Rev. Lett.} \textbf{1996}, \emph{77},
  3865--3868\relax
\mciteBstWouldAddEndPuncttrue
\mciteSetBstMidEndSepPunct{\mcitedefaultmidpunct}
{\mcitedefaultendpunct}{\mcitedefaultseppunct}\relax
\EndOfBibitem
\bibitem[Tran and Blaha(2009)Tran, and Blaha]{Tran:2009kk}
Tran,~F.; Blaha,~P. Accurate {Band} {Gaps} of {Semiconductors} and {Insulators}
  with a {Semilocal} {Exchange}-{Correlation} {Potential}. \emph{Phys. Rev.
  Lett.} \textbf{2009}, \emph{102}, 226401\relax
\mciteBstWouldAddEndPuncttrue
\mciteSetBstMidEndSepPunct{\mcitedefaultmidpunct}
{\mcitedefaultendpunct}{\mcitedefaultseppunct}\relax
\EndOfBibitem
\bibitem[Popescu and Zunger(2010)Popescu, and Zunger]{Popescu:2010jd}
Popescu,~V.; Zunger,~A. Effective {Band} {Structure} of {Random} {Alloys}.
  \emph{Phys. Rev. Lett.} \textbf{2010}, \emph{104}, 236403\relax
\mciteBstWouldAddEndPuncttrue
\mciteSetBstMidEndSepPunct{\mcitedefaultmidpunct}
{\mcitedefaultendpunct}{\mcitedefaultseppunct}\relax
\EndOfBibitem
\bibitem[Rubel \latin{et~al.}(2014)Rubel, Bokhanchuk, Ahmed, and
  Assmann]{Rubel:2014fv}
Rubel,~O.; Bokhanchuk,~A.; Ahmed,~S.~J.; Assmann,~E. Unfolding the Band
  Structure of Disordered Solids: {From} Bound States to High-Mobility {Kane}
  Fermions. \emph{Phys. Rev. B} \textbf{2014}, \emph{90}, 115202\relax
\mciteBstWouldAddEndPuncttrue
\mciteSetBstMidEndSepPunct{\mcitedefaultmidpunct}
{\mcitedefaultendpunct}{\mcitedefaultseppunct}\relax
\EndOfBibitem
\bibitem[Fuhr \latin{et~al.}(2013)Fuhr, Ventura, and Barrio]{Fuhr:2013fo}
Fuhr,~J.~D.; Ventura,~C.~I.; Barrio,~R.~A. Formation of Non-Substitutional
  $\beta$-Sn Defects in Ge$_{1-x}$Sn$_x$ Alloys. \emph{Journal of Applied
  Physics} \textbf{2013}, \emph{114}, 193508\relax
\mciteBstWouldAddEndPuncttrue
\mciteSetBstMidEndSepPunct{\mcitedefaultmidpunct}
{\mcitedefaultendpunct}{\mcitedefaultseppunct}\relax
\EndOfBibitem
\bibitem[Marceau \latin{et~al.}(2015)Marceau, Ceguerra, Breen, Raabe, and
  Ringer]{Marceau:2015go}
Marceau,~R. K.~W.; Ceguerra,~A.~V.; Breen,~A.~J.; Raabe,~D.; Ringer,~S.~P.
  {Quantitative chemical-structure evaluation using atom probe tomography:
  Short-range order analysis of Fe{\textendash}Al}. \emph{Ultramicroscopy}
  \textbf{2015}, \emph{157}, 12--20\relax
\mciteBstWouldAddEndPuncttrue
\mciteSetBstMidEndSepPunct{\mcitedefaultmidpunct}
{\mcitedefaultendpunct}{\mcitedefaultseppunct}\relax
\EndOfBibitem
\bibitem[Liu()]{Liu:2020}
To be submitted for publication\relax
\mciteBstWouldAddEndPuncttrue
\mciteSetBstMidEndSepPunct{\mcitedefaultmidpunct}
{\mcitedefaultendpunct}{\mcitedefaultseppunct}\relax
\EndOfBibitem
\bibitem[Gallagher \latin{et~al.}(2014)Gallagher, Senaratne, Kouvetakis, and
  Menéndez]{Gallagher:2014jj}
Gallagher,~J.~D.; Senaratne,~C.~L.; Kouvetakis,~J.; Menéndez,~J. Compositional
  Dependence of the Bowing Parameter for the Direct and Indirect Band Gaps in
  Ge$_{1-y}$Sn$_y$ Alloys. \emph{Appl. Phys. Lett.} \textbf{2014}, \emph{105},
  142102\relax
\mciteBstWouldAddEndPuncttrue
\mciteSetBstMidEndSepPunct{\mcitedefaultmidpunct}
{\mcitedefaultendpunct}{\mcitedefaultseppunct}\relax
\EndOfBibitem
\end{mcitethebibliography}

\providecommand{\latin}[1]{#1}
\providecommand*\mcitethebibliography{\thebibliography}
\csname @ifundefined\endcsname{endmcitethebibliography}
  {\let\endmcitethebibliography\endthebibliography}{}


\end{document}